\documentclass[reprint,amsmath,amssymb,aps,]{revtex4-1}
\usepackage[mathlines]{lineno}% Enable numbering of text and display math
\usepackage{hyperref}
\usepackage{amsmath,graphicx}
\usepackage{subcaption}

\begin{document}

\preprint{APS/123-QED}

\title{ADMX SLIC: Results from a Superconducting LC Circuit Investigating Cold Axions}

 \author{N. Crisosto}
 \email[Correspondence to:]{nmc25@uw.edu}
  \affiliation{University of Florida, Gainesville, FL 32611, USA}
  \affiliation{presently at University of Washington}
 \author{P. Sikivie}
  \affiliation{University of Florida, Gainesville, FL 32611, USA}
\author{N. S. Sullivan}
  \affiliation{University of Florida, Gainesville, FL 32611, USA}
\author{D. B. Tanner}
  \affiliation{University of Florida, Gainesville, FL 32611, USA}
  
      \author{J. Yang}%
  \affiliation{University of Washington, Seattle, WA 98195, USA}
    \author{G. Rybka}%
   
  \affiliation{University of Washington, Seattle, WA 98195, USA}
 \collaboration{ADMX Collaboration}%\noaffiliation
   \thanks{All of the authors also are members of the ADMX collaboration and the work reported reflects results obtained by those authors who worked directly on this project related to but distinct from ADMX.}

\date{\today}% It is always \today, today,

\begin{abstract}
Axions are a promising cold dark matter candidate.  Haloscopes, which use the conversion of axions to photons in the presence of a magnetic field to detect axions, are the basis of microwave cavity searches such as the Axion Dark Matter eXperiment (ADMX).  To search for lighter, low frequency axions in the sub $2\times10^{-7}$ eV  (50 MHz) range, a tunable lumped-element LC circuit has been proposed.   For the first time, through ADMX SLIC (\textbf{S}uperconducting \textbf{L}C Circuit \textbf{I}nvestigating \textbf{C}old Axions),  a resonant LC circuit was used to probe this region of axion mass-coupling space.  The detector used a superconducting LC circuit with piezoelectric driven capacitive tuning. The axion mass and corresponding frequency range $1.7498 -1.7519 \times10^{-7}$ eV (42.31 -- 42.36 MHz), $1.7734 -  1.7738 \times10^{-7}$ eV (42.88 -- 42.89 MHz), and $1.8007 - 1.8015 \times10^{-7}$ eV (43.54 -- 43.56 MHz) was covered at magnetic fields of 4.5 T, 5.0 T, and 7.0 T respectively.  Exclusion results from the search data, for coupling below $10^{-12}$ GeV $\textsuperscript{-1}$  are presented.  
%and and coupling below $10^{-10}$ GeV $\textsuperscript{-1}$,
\end{abstract}

\pacs{Valid PACS appear here}% PACS, the Physics and Astronomy
\maketitle

%\tableofcontents

%\linenumbers

%\section{Introduction}
 
The constituents of the dark matter of our Universe are yet to be accounted for.  Axions are a well-motivated dark matter candidate as they arise independently from the Peccei-Quinn solution to the strong CP Problem \cite{Peccei:1977hh,Weinberg:1977ma,Wilczek:1977pj}. If the axion mass is
in the $10^{-6}$ to $10^{-5}$ eV range, therefore very long-lived and very weakly coupled, then the cavity haloscope  \cite{Sikivie:1983ip,Sikivie1985} appears currently the best detection method.  The scheme is based on the electromagnetic coupling of axions to two photons: 
\begin{equation}
\mathcal{L}_{ a \gamma \gamma } = - ga(x) \vec{E(x)} \cdot \vec{B(x)} 
\end{equation}
where $g$ is a coupling constant, $a$ is the axion field,  $\vec{E}$ is the electric field, and $\vec{B}$ is the magnetic field.
In the presence of a strong magnetic field, an axion may convert into a real, detectable photon. A tuned resonator can subsequently enhance detection of the axion-sourced photon signal. The axion mass is unknown and the detector must be tuned through a range of possible axion masses. %PS
%The axion mass is a unknown and the detector must be tuned through the %range of possible axion-sourced frequencies corresponding to the axion mass. 
The frequency of the axion sourced photon signal, $\omega$, is set by the condition $\hbar \omega \approx m_a c^2 + \frac{1}{2} m_a v^2$, where $m_a$ is the axion mass and $v$ is axion velocity. 
The KSVZ \cite{Zhitnitsky:1980tq} and DFSZ \cite{DINE1983137} models are typically used to set $g_{\gamma}$, where $g = g_{\gamma} \frac{\alpha}{\pi f_{a}}$ and $f_{a}$ is the axion decay constant. 

Microwave cavity searches, including RBF \cite{DePanfilis:1987dk}, UF \cite{Hagmann:1990tj}, ADMX \cite{1538-4357-571-1-L27, Asztalos:2009yp, SLOAN201695, PhysRevD.94.082001, PhysRevLett.120.151301}, and HAYSTAC \cite{ PhysRevD.97.092001} have already scanned sections of axion parameter space.  To complement excluded axion mass parameter space, and probe couplings weaker than past helioscope searches \cite{cast2017}, an LC circuit  resonant structure in a lumped element regime can be used instead of a microwave cavity \cite{PhysRevLett.112.131301, unp00s, Cabrera}, and was done in the pilot experiment, ADMX SLIC, presented here.  This strategy in resonant structure, essentially a new class of contralto haloscopes, is also being pursued by  ABRACADABRA \cite{PhysRevLett.122.121802}, BEAST \cite{beast_2018},  and the Dark Matter Radio Experiment \cite{7750582}.  An optically pumped magnetometer readout system has also been proposed \cite{PhysRevD.97.072011}.  
 
% \section{Detection Strategy}

If the axion exists, Maxwell's equations are modified to become
\begin{equation}
\nabla \cdot \vec{E} = g \vec{B} \cdot \nabla a 
\end{equation}
\begin{equation}
\vec{\nabla} \times \vec{B} - \frac{\partial \vec{E} }{\partial t} = g (\vec{E} \times \vec{\nabla} a - \vec{B}\frac{\partial a}{\partial t}). 
\end{equation}
In the presence of a strong static magnetic field $B_0$, there is an axion-sourced current
\begin{equation}
\vec{j_{a}} = -g\vec{B_{0}}\frac{\partial a}{\partial t}.
\end{equation}
%where $\frac{\partial a}{\partial t} = -i \omega a$.  
Thus, there is a detectable oscillating magnetic field
\begin{equation}
\vec{\nabla} \times \vec{B_{a}} = \vec{j_{a}}
\end{equation}
in the quasi-static limit, when the length scale over with the external field $\vec{B_{0}}$ extends is much smaller than $c/ \omega$.
A loop antenna is used to capture the resulting magnetic flux
\begin{equation}
\Phi_{a} = -V_{m}g \frac{\partial a}{\partial t}B_{0}
\end{equation}
where $ V_{m} = \frac{1}{4}lr^2$ is set by the geometry of the loop antenna with height $l$ and width $r$.  Using the definition of inductance it follows that
\begin{equation}
I_{a}  = -\frac{\Phi_{a}}{L_1}
\end{equation}
in the limit of infinite capacitance.  
Including the resonant enhancement results in
\begin{equation}
I_{a} = \frac{Q}{L_{1}} V_{m} g \frac{\partial a}{\partial t} \vec{B_0}
\end{equation}
where $Q$ is the quality factor of the circuit.
In the case of the experiment described here, $I_{a}$ is inductively coupled to the input of a first-stage amplifier.  Flux in the coupling probe from mutual inductance, $M$, is given by
%%\begin{equation}
%V_{FET} = \alpha \sqrt{L_{1} L_{2}} \omega I_{a}.
%\end{equation}
\begin{equation}
\Phi_{p} = M I_{a} = \kappa \sqrt{L_{1} L_{2}} I_{a}
\end{equation}
where $L_{1}$ is the total inductance of the loop antenna and environment parasitics; $L_2$ is the coupling probe inductance, giving a signal power of
\begin{equation}
P_{input} =  \omega \kappa^2 L_{1} I_{a}^2
\end{equation}
where $\kappa$ is a coupling constant between $L_1$ and $L_2$.
The pilot experiment ADMX SLIC probes a lower-frequency, lighter-axion mass parameter space that is otherwise difficult to reach with existing microwave-cavity axion haloscope searches.  Prototype optimization studies led to a NbTi loop antenna capacitively tuned by a piezoelectric-actuated-dielectric deployed at 4.2 K in magnetic fields of 4.5--7 T.

%\section{Experiment Details}

%Description of how the LC search works, axion sourced B field that is being detected...
%Construction of LC experiement?
The loop antenna inductor of the LC circuit used in our axion search was a single rectangular loop, 7.62 cm $\times$ 31.25 cm, with copper-matrix-free 0.25 mm diameter NbTi wire strung around a polytetrafluoroethylene (PTFE) frame. A parallel capacitor was made with 5.08 cm $\times$ 5.72 cm NbTi plates, supported by PTFE blocks, and positioned with a 0.464 cm gap between plates.  PTFE screws fastened the form and secured the capacitor plates in place.  Angle bracket shapes were used to reduce weight and increase stability. A groove in the PTFE was used to secure most of the NbTi wire; PTFE tape secured the horizontal runs.  The NbTi capacitor plates were spot welded to the ends of the NbTi loop.  Tuning was achieved by moving an alumina sheet between the parallel plate capacitor with a rotary piezoelectric motor (ANR240). The calculated capacitance was 7.86 pF and 5.53 pF with and without the alumina fully inserted respectively.  Varactor tuning was previously explored here \cite{10.1007/978-3-319-92726-8_15, CrisostoThesis}, but was rejected due to observed high insertion loss.  Based on the loop antenna dimensions, the Terman formula \cite{TermanEquation} estimates an inductance of 0.968 $\mu$H.  Given the measured resonant frequency of 42 MHz with the capacitor tuned to its highest value, an effective total inductance, including environment parasitics, of 1.8 $\mu$H is found.  
The loop antenna is directly mounted to a $ \textsuperscript{3}$He refrigerator.  This placement allows for option of increased cooling in future runs and gives a representation of the resonator performance in situ of progressive cryogenic infrastructure.
%entire magnet picture
\begin{figure}[htpb]
  %\centering
    \includegraphics[width=3.4in]{{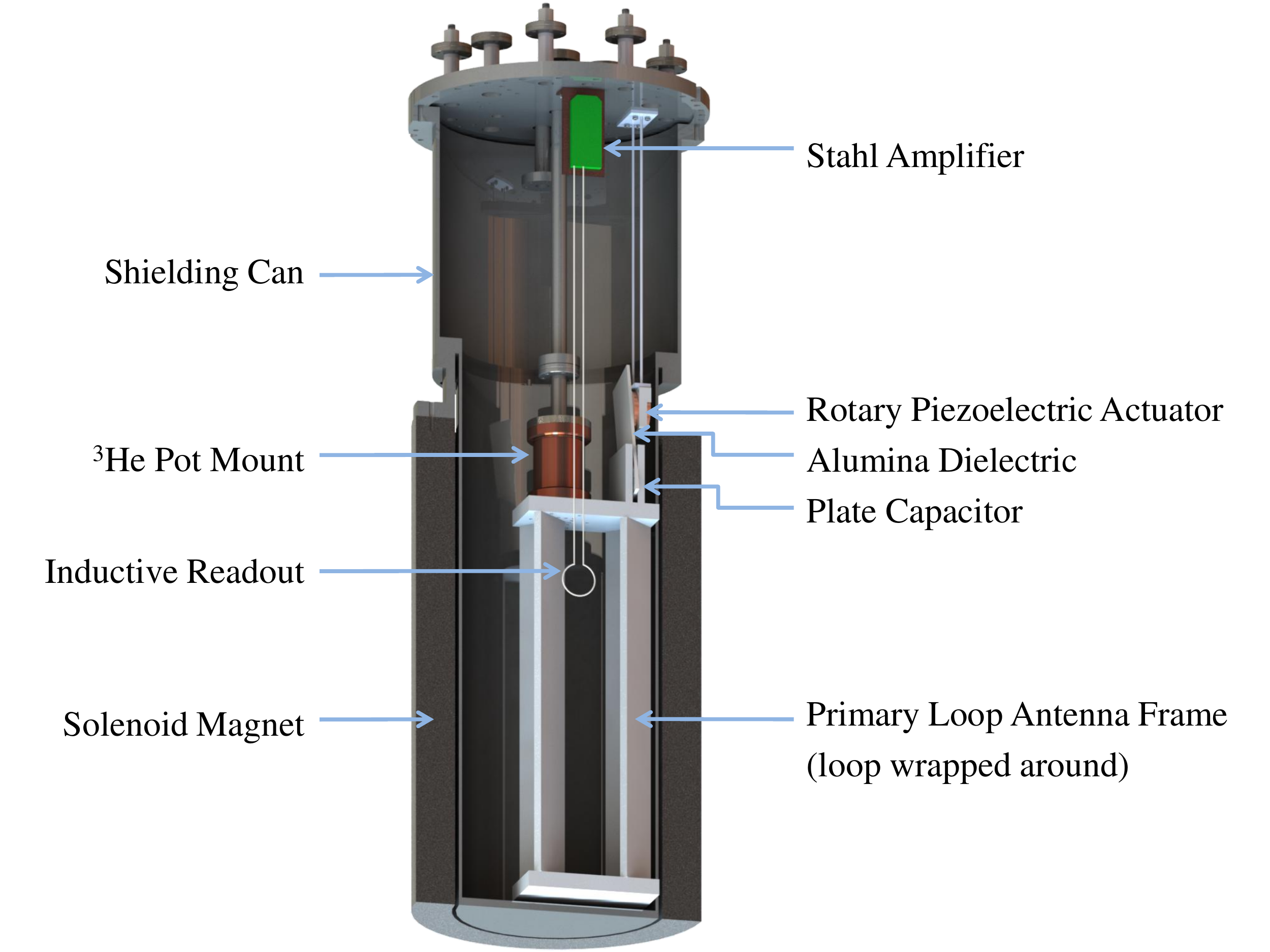}}
    \caption{A sketch of the experiment layout.  Two additional weakly coupled probes, for S$\textsubscript{21}$ measurements and optional amplifier bypass, are not shown.}
    \label{fig:Dim}
\end{figure}

%magnet
The magnet used is a Cryomagnetics superconducting (NbTi) solenoid with a 17.1 cm bore and a length of 40 cm. The central field is rated to 8.6 T at a current of 88 A, providing an average field of 7.5 T in a volume of about 8 liters.   %The loop antenna mounted before housed by the shielding can and magnet is show in in ~\ref{fig:LC_Full}.
In the course of data operations the vacuum can was not completely sealed to allow for cooling through immersion from $ \textsuperscript{4}$He.  The magnet is supported from a custom made stainless steel vacuum can.  % as shown in Figure ~\ref{fig:magnetmounted} .  
%shielding  
The entire insert is then housed in a super-insulating cryostat made by Precision Cryogenic System, Inc.  %add dimensions 
An an outline of the arrangement can be seen in Figure~\ref{fig:Dim}.  %expand smooth this out
The stainless steel vacuum can is lined with 0.003" Nb$\textsubscript{45}$Ti$\textsubscript{55}$ sheet for shielding of the loop antenna.

%pickup
A five turn NbTi, PTFE insulated coil was installed as an inductive coupling of the primary loop antenna to first-stage amplifier input.  NbTi leads of the inductive pickup were also connected to the amplifier input by crimping in CuNi capillary and soldering with PbSn. The highest $Q$ obtained, $Q >$ 10, 000, was in this configuration.  
%loop picture
%\begin{figure}[htpb]
%  \centering
%    \includegraphics[width=3in]{{images/LC_Full.jpg}}
%    \caption{Final LC circuit \textit{insitu} mounted to the $ \textsuperscript{3}$He refrigerator before installing the shielding can and magnet}
%    \label{fig:LC_Full}
%\end{figure}

%shield picture?

\begin{figure}[htpb]
  \centering
    \includegraphics[width=3.4in]{{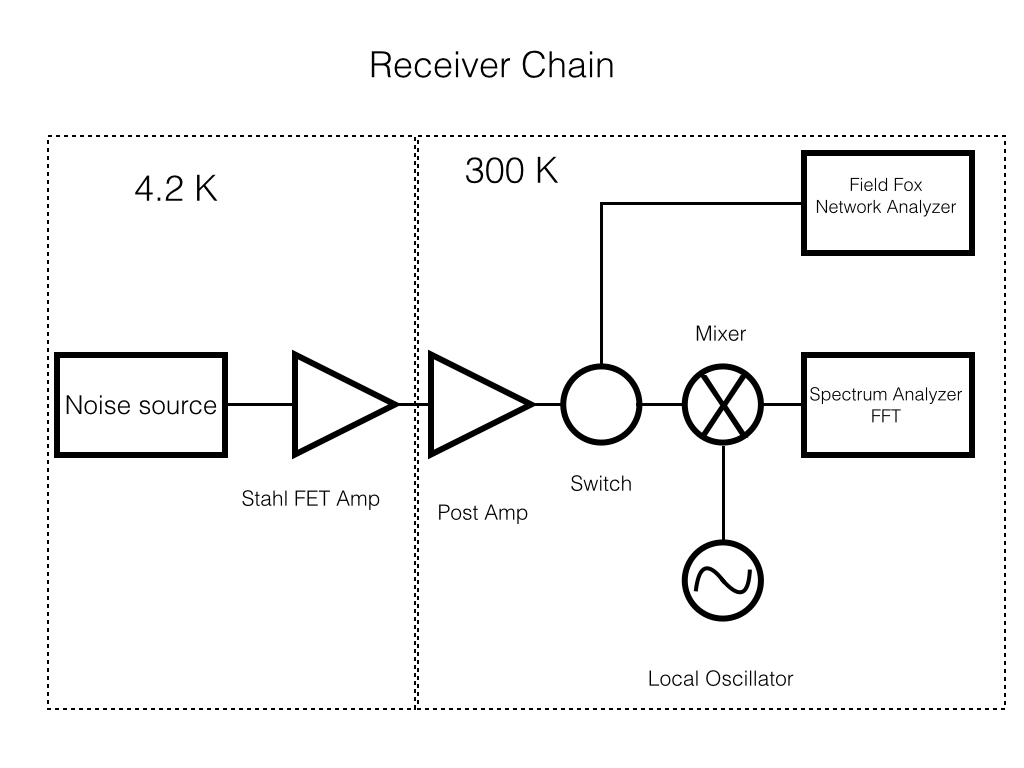}}
    \caption{Receiver chain diagram }
    \label{fig:DAQ}
\end{figure}

%entire magnet picture
%\begin{figure}[htpb]
%  \centering
%    \includegraphics[width=3.5in]{{images/magnet_loaded.jpg}}
%    \caption{Magnet loaded around shielding can before loading outer dewar }
%    \label{fig:magnetmounted}
%\end{figure}
%cooling to 4.2 K by 
%cryo amp
A low noise cryogenic GaAs FET amplifier \footnote{Stahl HDC-50} is used as the first-stage amplifier.
%room temp electronics
The output of the first-stage cryogenic amplifier above the vacuum can lid is connected via tinned copper braid PTFE insulated coaxial cable to room temperature \footnote{Pasternak PE-SR405FL}. After post amplification, the signal is fed to a Mini-Circuits ZX05-1-S double balanced mixer 
\footnote{Minicircuits ZX05-1+}
  and heterodyned to 10 kHz. An HP 856A is used as the local oscillator in mixing, controlled by GPIB and LabView drivers. The resulting
IF is read by an FFT SR760 Spectrum Analyzer and written to disk. 
In between tuning and data taking a Field Fox Microwave Analyzer N9916A is used to measure S-parameters.  A diagram of the receiver chain is shown in Figure~\ref{fig:DAQ}.
%add noise temp formulas and table and noise temp figure....?
From the Friis equation \cite{friis1944}, the noise temperature of the experiment is calculated to be 19.7 K \cite{CrisostoThesis}.  The background spectrum analyzer levels corresponded to Johnson voltage noise fluctuations at 23 K.  The greater of the two noise temperature figures is used in the reported analysis.  

\begin{figure}[htpb]
  \centering
    \includegraphics[width=3.4in]{{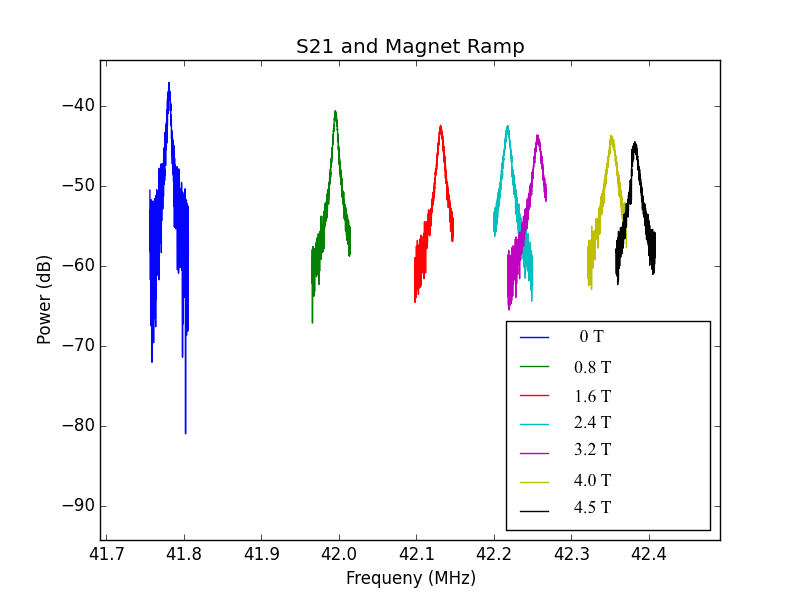}}
    \caption{S$\textsubscript{21}$ measurements of the loop antenna taken over the course of a magnet ramp.  A frequency shift and $Q$ reduction tracks the increased magnetic field.  }
    \label{fig:BQ}
\end{figure}

%performance
A reduction of $Q$ and frequency shift was found to track the magnet field ramp, as shown in Figure~\ref{fig:BQ}.   While ramping to 4.5 T, the $Q$ was reduced to $\sim$4,500.  At the highest magnetic fields in data operations, $Q$'s were typically 2,200-3,200.  Similar in field behavior has been reported in other superconducting resonators \cite{doi:10.1063/1.3271537, doi:10.1063/1.4958647}. %1T and 6T respectively.  
%magnetic field affect on Q and resonace

%\section{Data Operations}

%cascade plot???

%Data taking

Data operations consisted of tuning the LC circuit through its full range.  At each frequency step, signal is single heterodyne mixed down to 10 kHz and  a  FFT spectrum is then taken and digitized with a Stanford Research Systems 760 spectrum analyzer (SRS760).  An ANR240 rotary piezoelectric motor actuates a mechanical capacitor for the frequency scanning.  Between each tuning step quality factor, resonant frequency, and temperature are measured and recorded.  

The FFT spectra were taken with 12.5 kHz spans and 31.25 Hz wide bins. The SRS760 used takes 400 points per span and has a real time bandwidth of 100 kHz. Typically 10,000 averages of 32 ms scans were taken at each frequency and 1 kHz tuning steps were taken between bouts of spectra collection.  A tuning range of 42.3 - 50.0 MHz was expected but in operations was considerably smaller, probably from difficulties with the piezoelectric actuator when cold.
%Total acquisition time = averages * nyquist sampling rate limit
Intermittently, synthetic signals were injected, on a weakly coupled probe, and observed to verify data taking operations.
%inject on--> weak, did both white noise and sinuisoid from a function generator]%
Run 1 collected data at 4.5 T from 6/16/2018--6/19/2018.  Run 2 collected data at 5 T and 7 T from 7/19/2018--7/27/2018. 

%\section{Analysis}

The signal-to-noise ratio and expected axion sourced power are used to place limits on the coupling of the axion to two photons from the measured power spectra.  The signal-to-noise is calculated from the Dicke radiometer equation \cite{doi:10.1063/1.1770483} using power on the input coil of the first-stage amplifier,
\begin{equation}
SNR = \frac{P_{input}}{kT}\sqrt{\frac{\Delta t}{\Delta b}}
\end{equation}
where k is Boltzman's constant,  $\Delta t$ is integration time, and $\Delta b$ is bandwidth.
For each raw spectrum 1 kHz was trimmed off the ends before being co-added to form a grand spectrum. Background subtraction was done through a Savitsky-Golay filter.  
A significant power excess above average noise power could be indicative of an axion conversion signal \cite{Hagmann:1990tj} and a candidate.  A power excess above Gaussian background noise can be calculated from
\begin{equation}
\# \sigma = \frac{P - \bar P}{\bar P} {\sqrt{N}}.
\end{equation}

Numerous peaks of power excess were observed in the raw spectra, which are likely radio frequency interference (RFI) from external sources, but were not re-scanned as part of this study.  Axion couplings that would produce signals larger than the observed excesses can be excluded, as shown in the blue lines of Fig. 4.  Future large scale experiments will be able to implement more shielding and re-scan excesses to distinguish axion signals from RFI; for the purposes of understanding scaling of these experiments, the estimated limits that could have been achieved were RFI candidates eliminated are shown in orange.  It is interesting to note that even limited by RFI, unexplored axion-like-particle dark matter parameter space can be excluded.
\begin{figure*}[htb!]
    
      \begin{subfigure}[t]{0.52\textwidth}
    \includegraphics[width=\textwidth]{{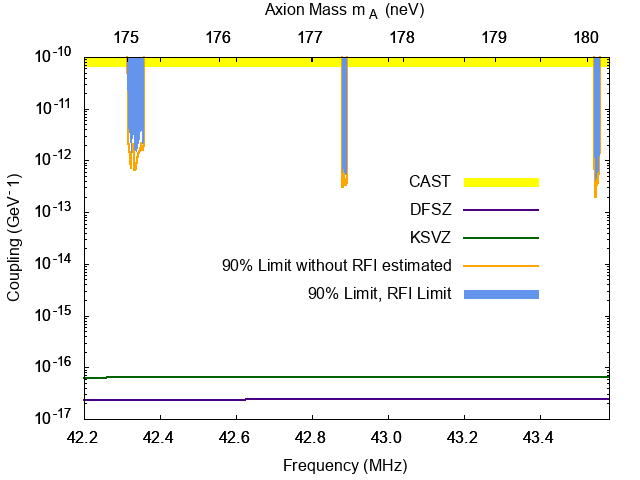}}
    \caption{Limit plot from all combined data runs}
    \label{fig:limit}
  \end{subfigure}

  \begin{subfigure}[t]{0.49\textwidth}
    \includegraphics[width=\textwidth]{{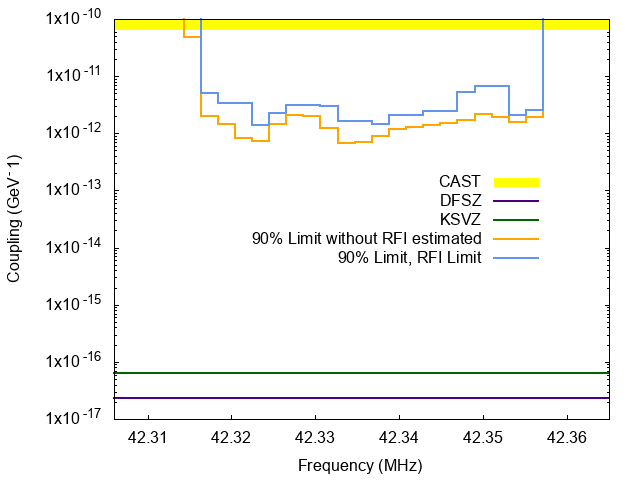}}
    \caption{Run 1 with a 4.5 T magnetic field}
     	 	\label{fig:limit_1}
  \end{subfigure}
~
  \begin{subfigure}[t]{0.49\textwidth}
    \includegraphics[width=\textwidth]{{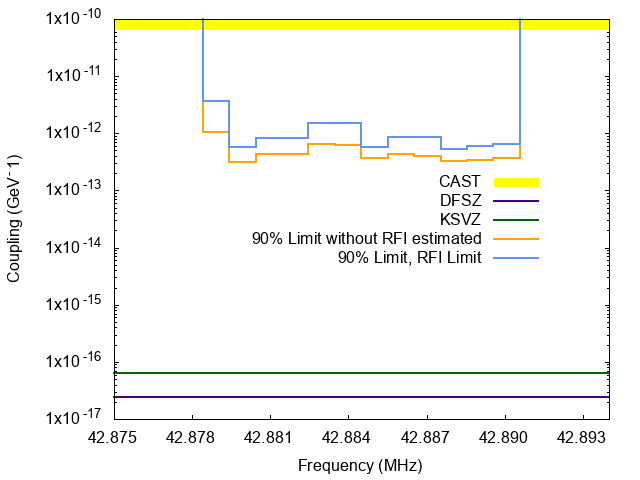}}
    \caption{Run 2.a with a 5.0 T magnetic field}
    \label{fig:limit_2a}
  \end{subfigure}
    \begin{subfigure}[t]{0.5\textwidth}
    \includegraphics[width=\textwidth]{{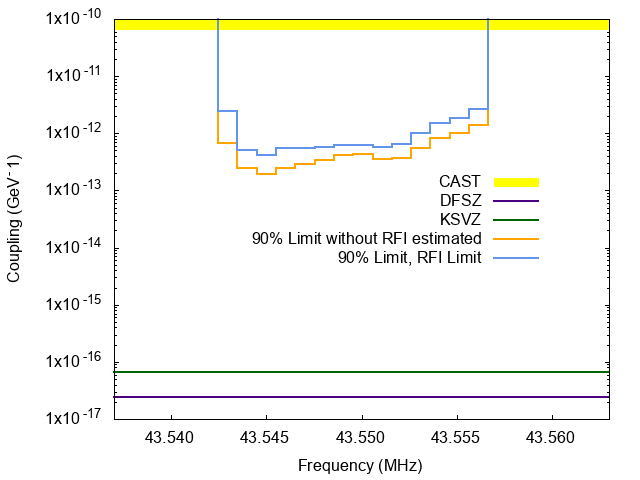}}
    \caption{Run 2.b with a 7.0 T magnetic field}
    \label{fig:limit_2b}
  \end{subfigure}
\caption{Individual and combined limits collected by ADMX SLIC}
  \label{fig:limit_plots}
\end{figure*}
\clearpage

Figure~\ref{fig:limit_plots} shows the axion-photon coupling exclusions achieved in the experiment. Panels \ref{fig:limit_1}, \ref{fig:limit_2a}, and \ref{fig:limit_2b} show the individual runs while panel \ref{fig:limit} shows the regions combined. The previous best limits set by CAST are also shown.  The limits presented rule out axions with couplings greater than the limit-line.  
%what noise temperature is this equivalent to?
%show histogram to show noisiness????
%limits plots
%\begin{figure}[htpb]
%limits in light of cast and abracadabra

%even more big picture limits

%\clearpage %current attempt to get conclusion on its own page and limts all on own page preceding
%\section{Conclusions}
An LC circuit based axion search presents a promising approach to scan unexplored regions of low-frequency axion parameter space.  Extensive prototype testing culminated in a piezoelectric-actuated dielectric-tuned NbTi superconducting LC circuit.  The operation of this new form of axion detector tested design considerations for future implementations including the measurements of AC electrical losses in superconducting circuits in a large magnetic field.  Our pilot experiment operated in 4.2 K and magnetic fields of 4.5--7 T.  The axion mass range of 
$1.7498 -1.7519 \times10^{-7}$ eV (42.31 -- 42.36 MHz), $1.7734 -  1.7738 \times10^{-7}$ eV (42.88 -- 42.89 MHz), and $1.8007 - 1.8015 \times10^{-7}$ eV (43.54 -- 43.56 MHz)
%$1.749-1.752 \times10^{-7}$ eV (42.31--42.36 MHz), $1.7734-1.7738 \times10^{-7}$ eV (42.88--42.89 MHz), and $1.8007-1.8015  \times10^{-7}$ eV (43.54--43.56 MHz) 
was searched.  A large amount of external RF noise was found in the data.  Despite an external RFI noise limited analysis a new section of axion mass and coupling was excluded. 

%\section{Acknowledgments}
%Updated grant list 

Research at the University of Florida has been supported by the US Department of Energy through grants DE-SC0010296 and DE-SC0009723TDD. Additional work at the University of Washington is supported by the US Department of Energy through grant DE-SC0011665. The authors gratefully acknowledge useful comments by Gianpaolo Carosi, John Clarke, Joe Gleason, Greg Labbe, Bill Malphurs, and Leslie Rosenberg.  The authors also gratefully acknowledge the larger ADMX collaboration beyond those responsible for this particular work who provided comments on the draft.

\bibliographystyle{apsrev4-1}
\bibliography{main}

\begin{thebibliography}{33}%
\makeatletter
\providecommand \@ifxundefined [1]{%
 \@ifx{#1\undefined}
}%
\providecommand \@ifnum [1]{%
 \ifnum #1\expandafter \@firstoftwo
 \else \expandafter \@secondoftwo
 \fi
}%
\providecommand \@ifx [1]{%
 \ifx #1\expandafter \@firstoftwo
 \else \expandafter \@secondoftwo
 \fi
}%
\providecommand \natexlab [1]{#1}%
\providecommand \enquote  [1]{``#1''}%
\providecommand \bibnamefont  [1]{#1}%
\providecommand \bibfnamefont [1]{#1}%
\providecommand \citenamefont [1]{#1}%
\providecommand \href@noop [0]{\@secondoftwo}%
\providecommand \href [0]{\begingroup \@sanitize@url \@href}%
\providecommand \@href[1]{\@@startlink{#1}\@@href}%
\providecommand \@@href[1]{\endgroup#1\@@endlink}%
\providecommand \@sanitize@url [0]{\catcode `\\12\catcode `\$12\catcode
  `\&12\catcode `\#12\catcode `\^12\catcode `\_12\catcode `\%12\relax}%
\providecommand \@@startlink[1]{}%
\providecommand \@@endlink[0]{}%
\providecommand \url  [0]{\begingroup\@sanitize@url \@url }%
\providecommand \@url [1]{\endgroup\@href {#1}{\urlprefix }}%
\providecommand \urlprefix  [0]{URL }%
\providecommand \Eprint [0]{\href }%
\providecommand \doibase [0]{http://dx.doi.org/}%
\providecommand \selectlanguage [0]{\@gobble}%
\providecommand \bibinfo  [0]{\@secondoftwo}%
\providecommand \bibfield  [0]{\@secondoftwo}%
\providecommand \translation [1]{[#1]}%
\providecommand \BibitemOpen [0]{}%
\providecommand \bibitemStop [0]{}%
\providecommand \bibitemNoStop [0]{.\EOS\space}%
\providecommand \EOS [0]{\spacefactor3000\relax}%
\providecommand \BibitemShut  [1]{\csname bibitem#1\endcsname}%
\let\auto@bib@innerbib\@empty
%</preamble>
\bibitem [{\citenamefont {Peccei}\ and\ \citenamefont
  {Quinn}(1977)}]{Peccei:1977hh}%
  \BibitemOpen
  \bibfield  {author} {\bibinfo {author} {\bibfnamefont {R.~D.}\ \bibnamefont
  {Peccei}}\ and\ \bibinfo {author} {\bibfnamefont {H.~R.}\ \bibnamefont
  {Quinn}},\ }\href {\doibase 10.1103/PhysRevLett.38.1440} {\bibfield
  {journal} {\bibinfo  {journal} {Phys. Rev. Lett.}\ }\textbf {\bibinfo
  {volume} {38}},\ \bibinfo {pages} {1440} (\bibinfo {year}
  {1977})}\BibitemShut {NoStop}%
%%CITATION = PRLTA,38,1440;%%
\bibitem [{\citenamefont {Weinberg}(1978)}]{Weinberg:1977ma}%
  \BibitemOpen
  \bibfield  {author} {\bibinfo {author} {\bibfnamefont {S.}~\bibnamefont
  {Weinberg}},\ }\href {\doibase 10.1103/PhysRevLett.40.223} {\bibfield
  {journal} {\bibinfo  {journal} {Phys. Rev. Lett.}\ }\textbf {\bibinfo
  {volume} {40}},\ \bibinfo {pages} {223} (\bibinfo {year} {1978})}\BibitemShut
  {NoStop}%
%%CITATION = PRLTA,40,223;%%
\bibitem [{\citenamefont {Wilczek}(1978)}]{Wilczek:1977pj}%
  \BibitemOpen
  \bibfield  {author} {\bibinfo {author} {\bibfnamefont {F.}~\bibnamefont
  {Wilczek}},\ }\href {\doibase 10.1103/PhysRevLett.40.279} {\bibfield
  {journal} {\bibinfo  {journal} {Phys. Rev. Lett.}\ }\textbf {\bibinfo
  {volume} {40}},\ \bibinfo {pages} {279} (\bibinfo {year} {1978})}\BibitemShut
  {NoStop}%
%%CITATION = PRLTA,40,279;%%
\bibitem [{\citenamefont {Sikivie}(1983)}]{Sikivie:1983ip}%
  \BibitemOpen
  \bibfield  {author} {\bibinfo {author} {\bibfnamefont {P.}~\bibnamefont
  {Sikivie}},\ }\href {\doibase 10.1103/PhysRevLett.51.1415} {\bibfield
  {journal} {\bibinfo  {journal} {Phys. Rev. Lett.}\ }\textbf {\bibinfo
  {volume} {51}},\ \bibinfo {pages} {1415} (\bibinfo {year}
  {1983})}\BibitemShut {NoStop}%
%%CITATION = PRLTA,51,1415;%%
\bibitem [{\citenamefont {Sikivie}(1985)}]{Sikivie1985}%
  \BibitemOpen
  \bibfield  {author} {\bibinfo {author} {\bibfnamefont {P.}~\bibnamefont
  {Sikivie}},\ }\href {\doibase 10.1103/PhysRevD.32.2988} {\bibfield  {journal}
  {\bibinfo  {journal} {Phys. Rev. D}\ }\textbf {\bibinfo {volume} {32}},\
  \bibinfo {pages} {2988} (\bibinfo {year} {1985})}\BibitemShut {NoStop}%
\bibitem [{\citenamefont {Zhitnitsky}(1980)}]{Zhitnitsky:1980tq}%
  \BibitemOpen
  \bibfield  {author} {\bibinfo {author} {\bibfnamefont {A.}~\bibnamefont
  {Zhitnitsky}},\ }\href@noop {} {\bibfield  {journal} {\bibinfo  {journal}
  {Sov.J.Nucl.Phys.}\ }\textbf {\bibinfo {volume} {31}},\ \bibinfo {pages}
  {260} (\bibinfo {year} {1980})}\BibitemShut {NoStop}%
%%CITATION = SJNCA,31,260;%%
\bibitem [{\citenamefont {Dine}\ and\ \citenamefont
  {Fischler}(1983)}]{DINE1983137}%
  \BibitemOpen
  \bibfield  {author} {\bibinfo {author} {\bibfnamefont {M.}~\bibnamefont
  {Dine}}\ and\ \bibinfo {author} {\bibfnamefont {W.}~\bibnamefont
  {Fischler}},\ }\href {\doibase https://doi.org/10.1016/0370-2693(83)90639-1}
  {\bibfield  {journal} {\bibinfo  {journal} {Phys. Lett. B}\ }\textbf
  {\bibinfo {volume} {120}},\ \bibinfo {pages} {137 } (\bibinfo {year}
  {1983})}\BibitemShut {NoStop}%
\bibitem [{\citenamefont {De~Panfilis}\ \emph {et~al.}(1987)\citenamefont
  {De~Panfilis}, \citenamefont {Melissinos}, \citenamefont {Moskowitz},
  \citenamefont {Rogers}, \citenamefont {Semertzidis} \emph
  {et~al.}}]{DePanfilis:1987dk}%
  \BibitemOpen
  \bibfield  {author} {\bibinfo {author} {\bibfnamefont {S.}~\bibnamefont
  {De~Panfilis}}, \bibinfo {author} {\bibfnamefont {A.}~\bibnamefont
  {Melissinos}}, \bibinfo {author} {\bibfnamefont {B.}~\bibnamefont
  {Moskowitz}}, \bibinfo {author} {\bibfnamefont {J.}~\bibnamefont {Rogers}},
  \bibinfo {author} {\bibfnamefont {Y.}~\bibnamefont {Semertzidis}},  \emph
  {et~al.},\ }\href {\doibase 10.1103/PhysRevLett.59.839} {\bibfield  {journal}
  {\bibinfo  {journal} {Phys. Rev. Lett.}\ }\textbf {\bibinfo {volume} {59}},\
  \bibinfo {pages} {839} (\bibinfo {year} {1987})}\BibitemShut {NoStop}%
%%CITATION = PRLTA,59,839;%%
\bibitem [{\citenamefont {Hagmann}\ \emph {et~al.}(1990)\citenamefont
  {Hagmann}, \citenamefont {Sikivie}, \citenamefont {Sullivan},\ and\
  \citenamefont {Tanner}}]{Hagmann:1990tj}%
  \BibitemOpen
  \bibfield  {author} {\bibinfo {author} {\bibfnamefont {C.}~\bibnamefont
  {Hagmann}}, \bibinfo {author} {\bibfnamefont {P.}~\bibnamefont {Sikivie}},
  \bibinfo {author} {\bibfnamefont {N.~S.}\ \bibnamefont {Sullivan}}, \ and\
  \bibinfo {author} {\bibfnamefont {D.~B.}\ \bibnamefont {Tanner}},\ }\href
  {\doibase 10.1103/PhysRevD.42.1297} {\bibfield  {journal} {\bibinfo
  {journal} {Phys. Rev.}\ }\textbf {\bibinfo {volume} {D42}},\ \bibinfo {pages}
  {1297} (\bibinfo {year} {1990})}\BibitemShut {NoStop}%
%%CITATION = PHRVA,D42,1297;%%
\bibitem [{\citenamefont {Asztalos}\ \emph {et~al.}(2002)\citenamefont
  {Asztalos}, \citenamefont {Daw}, \citenamefont {Peng}, \citenamefont
  {Rosenberg}, \citenamefont {Yu}, \citenamefont {Hagmann}, \citenamefont
  {Kinion}, \citenamefont {Stoeffl}, \citenamefont {van Bibber}, \citenamefont
  {LaVeigne}, \citenamefont {Sikivie}, \citenamefont {Sullivan}, \citenamefont
  {Tanner}, \citenamefont {Nezrick},\ and\ \citenamefont
  {Moltz}}]{1538-4357-571-1-L27}%
  \BibitemOpen
  \bibfield  {author} {\bibinfo {author} {\bibfnamefont {S.~J.}\ \bibnamefont
  {Asztalos}}, \bibinfo {author} {\bibfnamefont {E.}~\bibnamefont {Daw}},
  \bibinfo {author} {\bibfnamefont {H.}~\bibnamefont {Peng}}, \bibinfo {author}
  {\bibfnamefont {L.~J.}\ \bibnamefont {Rosenberg}}, \bibinfo {author}
  {\bibfnamefont {D.~B.}\ \bibnamefont {Yu}}, \bibinfo {author} {\bibfnamefont
  {C.}~\bibnamefont {Hagmann}}, \bibinfo {author} {\bibfnamefont
  {D.}~\bibnamefont {Kinion}}, \bibinfo {author} {\bibfnamefont
  {W.}~\bibnamefont {Stoeffl}}, \bibinfo {author} {\bibfnamefont
  {K.}~\bibnamefont {van Bibber}}, \bibinfo {author} {\bibfnamefont
  {J.}~\bibnamefont {LaVeigne}}, \bibinfo {author} {\bibfnamefont
  {P.}~\bibnamefont {Sikivie}}, \bibinfo {author} {\bibfnamefont {N.~S.}\
  \bibnamefont {Sullivan}}, \bibinfo {author} {\bibfnamefont {D.~B.}\
  \bibnamefont {Tanner}}, \bibinfo {author} {\bibfnamefont {F.}~\bibnamefont
  {Nezrick}}, \ and\ \bibinfo {author} {\bibfnamefont {D.~M.}\ \bibnamefont
  {Moltz}},\ }\href {http://stacks.iop.org/1538-4357/571/i=1/a=L27} {\bibfield
  {journal} {\bibinfo  {journal} {The Astrophysical Journal Letters}\ }\textbf
  {\bibinfo {volume} {571}},\ \bibinfo {pages} {L27} (\bibinfo {year}
  {2002})}\BibitemShut {NoStop}%
\bibitem [{\citenamefont {Asztalos}\ \emph {et~al.}(2010)\citenamefont
  {Asztalos}, \citenamefont {Carosi}, \citenamefont {Hagmann}, \citenamefont
  {Kinion}, \citenamefont {van Bibber}, \citenamefont {Hotz}, \citenamefont
  {Rosenberg}, \citenamefont {Rybka}, \citenamefont {Hoskins}, \citenamefont
  {Hwang}, \citenamefont {Sikivie}, \citenamefont {Tanner}, \citenamefont
  {Bradley},\ and\ \citenamefont {Clarke}}]{Asztalos:2009yp}%
  \BibitemOpen
  \bibfield  {author} {\bibinfo {author} {\bibfnamefont {S.~J.}\ \bibnamefont
  {Asztalos}}, \bibinfo {author} {\bibfnamefont {G.}~\bibnamefont {Carosi}},
  \bibinfo {author} {\bibfnamefont {C.}~\bibnamefont {Hagmann}}, \bibinfo
  {author} {\bibfnamefont {D.}~\bibnamefont {Kinion}}, \bibinfo {author}
  {\bibfnamefont {K.}~\bibnamefont {van Bibber}}, \bibinfo {author}
  {\bibfnamefont {M.}~\bibnamefont {Hotz}}, \bibinfo {author} {\bibfnamefont
  {L.~J.}\ \bibnamefont {Rosenberg}}, \bibinfo {author} {\bibfnamefont
  {G.}~\bibnamefont {Rybka}}, \bibinfo {author} {\bibfnamefont
  {J.}~\bibnamefont {Hoskins}}, \bibinfo {author} {\bibfnamefont
  {J.}~\bibnamefont {Hwang}}, \bibinfo {author} {\bibfnamefont
  {P.}~\bibnamefont {Sikivie}}, \bibinfo {author} {\bibfnamefont {D.~B.}\
  \bibnamefont {Tanner}}, \bibinfo {author} {\bibfnamefont {R.}~\bibnamefont
  {Bradley}}, \ and\ \bibinfo {author} {\bibfnamefont {J.}~\bibnamefont
  {Clarke}},\ }\href {\doibase 10.1103/PhysRevLett.104.041301} {\bibfield
  {journal} {\bibinfo  {journal} {Phys. Rev. Lett.}\ }\textbf {\bibinfo
  {volume} {104}},\ \bibinfo {pages} {041301} (\bibinfo {year}
  {2010})}\BibitemShut {NoStop}%
\bibitem [{\citenamefont {Sloan}\ \emph {et~al.}(2016)\citenamefont {Sloan},
  \citenamefont {Hotz}, \citenamefont {Boutan}, \citenamefont {Bradley},
  \citenamefont {Carosi}, \citenamefont {Carter}, \citenamefont {Clarke},
  \citenamefont {Crisosto}, \citenamefont {Daw}, \citenamefont {Gleason},
  \citenamefont {Hoskins}, \citenamefont {Khatiwada}, \citenamefont
  {Lyapustin}, \citenamefont {Malagon}, \citenamefont {O'Kelley}, \citenamefont
  {Ottens}, \citenamefont {Rosenberg}, \citenamefont {Rybka}, \citenamefont
  {Stern}, \citenamefont {Sullivan}, \citenamefont {Tanner}, \citenamefont {van
  Bibber}, \citenamefont {Wagner},\ and\ \citenamefont {Will}}]{SLOAN201695}%
  \BibitemOpen
  \bibfield  {author} {\bibinfo {author} {\bibfnamefont {J.}~\bibnamefont
  {Sloan}}, \bibinfo {author} {\bibfnamefont {M.}~\bibnamefont {Hotz}},
  \bibinfo {author} {\bibfnamefont {C.}~\bibnamefont {Boutan}}, \bibinfo
  {author} {\bibfnamefont {R.}~\bibnamefont {Bradley}}, \bibinfo {author}
  {\bibfnamefont {G.}~\bibnamefont {Carosi}}, \bibinfo {author} {\bibfnamefont
  {D.}~\bibnamefont {Carter}}, \bibinfo {author} {\bibfnamefont
  {J.}~\bibnamefont {Clarke}}, \bibinfo {author} {\bibfnamefont
  {N.}~\bibnamefont {Crisosto}}, \bibinfo {author} {\bibfnamefont
  {E.}~\bibnamefont {Daw}}, \bibinfo {author} {\bibfnamefont {J.}~\bibnamefont
  {Gleason}}, \bibinfo {author} {\bibfnamefont {J.}~\bibnamefont {Hoskins}},
  \bibinfo {author} {\bibfnamefont {R.}~\bibnamefont {Khatiwada}}, \bibinfo
  {author} {\bibfnamefont {D.}~\bibnamefont {Lyapustin}}, \bibinfo {author}
  {\bibfnamefont {A.}~\bibnamefont {Malagon}}, \bibinfo {author} {\bibfnamefont
  {S.}~\bibnamefont {O'Kelley}}, \bibinfo {author} {\bibfnamefont {R.~S.}\
  \bibnamefont {Ottens}}, \bibinfo {author} {\bibfnamefont {L.~J.}\
  \bibnamefont {Rosenberg}}, \bibinfo {author} {\bibfnamefont {G.}~\bibnamefont
  {Rybka}}, \bibinfo {author} {\bibfnamefont {I.}~\bibnamefont {Stern}},
  \bibinfo {author} {\bibfnamefont {N.~S.}\ \bibnamefont {Sullivan}}, \bibinfo
  {author} {\bibfnamefont {D.~B.}\ \bibnamefont {Tanner}}, \bibinfo {author}
  {\bibfnamefont {K.}~\bibnamefont {van Bibber}}, \bibinfo {author}
  {\bibfnamefont {A.}~\bibnamefont {Wagner}}, \ and\ \bibinfo {author}
  {\bibfnamefont {D.}~\bibnamefont {Will}},\ }\href {\doibase
  10.1016/j.dark.2016.09.003} {\bibfield  {journal} {\bibinfo  {journal}
  {Physics of the Dark Universe}\ }\textbf {\bibinfo {volume} {14}},\ \bibinfo
  {pages} {95 } (\bibinfo {year} {2016})}\BibitemShut {NoStop}%
\bibitem [{\citenamefont {Hoskins}\ \emph {et~al.}(2016)\citenamefont
  {Hoskins}, \citenamefont {Crisosto}, \citenamefont {Gleason}, \citenamefont
  {Sikivie}, \citenamefont {Stern}, \citenamefont {Sullivan}, \citenamefont
  {Tanner}, \citenamefont {Boutan}, \citenamefont {Hotz}, \citenamefont
  {Khatiwada}, \citenamefont {Lyapustin}, \citenamefont {Malagon},
  \citenamefont {Ottens}, \citenamefont {Rosenberg}, \citenamefont {Rybka},
  \citenamefont {Sloan}, \citenamefont {Wagner}, \citenamefont {Will},
  \citenamefont {Carosi}, \citenamefont {Carter}, \citenamefont {Duffy},
  \citenamefont {Bradley}, \citenamefont {Clarke}, \citenamefont {O'Kelley},
  \citenamefont {van Bibber},\ and\ \citenamefont {Daw}}]{PhysRevD.94.082001}%
  \BibitemOpen
  \bibfield  {author} {\bibinfo {author} {\bibfnamefont {J.}~\bibnamefont
  {Hoskins}}, \bibinfo {author} {\bibfnamefont {N.}~\bibnamefont {Crisosto}},
  \bibinfo {author} {\bibfnamefont {J.}~\bibnamefont {Gleason}}, \bibinfo
  {author} {\bibfnamefont {P.}~\bibnamefont {Sikivie}}, \bibinfo {author}
  {\bibfnamefont {I.}~\bibnamefont {Stern}}, \bibinfo {author} {\bibfnamefont
  {N.~S.}\ \bibnamefont {Sullivan}}, \bibinfo {author} {\bibfnamefont {D.~B.}\
  \bibnamefont {Tanner}}, \bibinfo {author} {\bibfnamefont {C.}~\bibnamefont
  {Boutan}}, \bibinfo {author} {\bibfnamefont {M.}~\bibnamefont {Hotz}},
  \bibinfo {author} {\bibfnamefont {R.}~\bibnamefont {Khatiwada}}, \bibinfo
  {author} {\bibfnamefont {D.}~\bibnamefont {Lyapustin}}, \bibinfo {author}
  {\bibfnamefont {A.}~\bibnamefont {Malagon}}, \bibinfo {author} {\bibfnamefont
  {R.}~\bibnamefont {Ottens}}, \bibinfo {author} {\bibfnamefont {L.~J.}\
  \bibnamefont {Rosenberg}}, \bibinfo {author} {\bibfnamefont {G.}~\bibnamefont
  {Rybka}}, \bibinfo {author} {\bibfnamefont {J.}~\bibnamefont {Sloan}},
  \bibinfo {author} {\bibfnamefont {A.}~\bibnamefont {Wagner}}, \bibinfo
  {author} {\bibfnamefont {D.}~\bibnamefont {Will}}, \bibinfo {author}
  {\bibfnamefont {G.}~\bibnamefont {Carosi}}, \bibinfo {author} {\bibfnamefont
  {D.}~\bibnamefont {Carter}}, \bibinfo {author} {\bibfnamefont {L.~D.}\
  \bibnamefont {Duffy}}, \bibinfo {author} {\bibfnamefont {R.}~\bibnamefont
  {Bradley}}, \bibinfo {author} {\bibfnamefont {J.}~\bibnamefont {Clarke}},
  \bibinfo {author} {\bibfnamefont {S.}~\bibnamefont {O'Kelley}}, \bibinfo
  {author} {\bibfnamefont {K.}~\bibnamefont {van Bibber}}, \ and\ \bibinfo
  {author} {\bibfnamefont {E.~J.}\ \bibnamefont {Daw}},\ }\href {\doibase
  10.1103/PhysRevD.94.082001} {\bibfield  {journal} {\bibinfo  {journal} {Phys.
  Rev. D}\ }\textbf {\bibinfo {volume} {94}},\ \bibinfo {pages} {082001}
  (\bibinfo {year} {2016})}\BibitemShut {NoStop}%
\bibitem [{\citenamefont {Du}\ \emph {et~al.}(2018)\citenamefont {Du},
  \citenamefont {Force}, \citenamefont {Khatiwada}, \citenamefont {Lentz},
  \citenamefont {Ottens}, \citenamefont {Rosenberg}, \citenamefont {Rybka},
  \citenamefont {Carosi}, \citenamefont {Woollett}, \citenamefont {Bowring},
  \citenamefont {Chou}, \citenamefont {Sonnenschein}, \citenamefont {Wester},
  \citenamefont {Boutan}, \citenamefont {Oblath}, \citenamefont {Bradley},
  \citenamefont {Daw}, \citenamefont {Dixit}, \citenamefont {Clarke},
  \citenamefont {O'Kelley}, \citenamefont {Crisosto}, \citenamefont {Gleason},
  \citenamefont {Jois}, \citenamefont {Sikivie}, \citenamefont {Stern},
  \citenamefont {Sullivan}, \citenamefont {Tanner},\ and\ \citenamefont
  {Hilton}}]{PhysRevLett.120.151301}%
  \BibitemOpen
  \bibfield  {author} {\bibinfo {author} {\bibfnamefont {N.}~\bibnamefont
  {Du}}, \bibinfo {author} {\bibfnamefont {N.}~\bibnamefont {Force}}, \bibinfo
  {author} {\bibfnamefont {R.}~\bibnamefont {Khatiwada}}, \bibinfo {author}
  {\bibfnamefont {E.}~\bibnamefont {Lentz}}, \bibinfo {author} {\bibfnamefont
  {R.}~\bibnamefont {Ottens}}, \bibinfo {author} {\bibfnamefont {L.~J.}\
  \bibnamefont {Rosenberg}}, \bibinfo {author} {\bibfnamefont {G.}~\bibnamefont
  {Rybka}}, \bibinfo {author} {\bibfnamefont {G.}~\bibnamefont {Carosi}},
  \bibinfo {author} {\bibfnamefont {N.}~\bibnamefont {Woollett}}, \bibinfo
  {author} {\bibfnamefont {D.}~\bibnamefont {Bowring}}, \bibinfo {author}
  {\bibfnamefont {A.~S.}\ \bibnamefont {Chou}}, \bibinfo {author}
  {\bibfnamefont {A.}~\bibnamefont {Sonnenschein}}, \bibinfo {author}
  {\bibfnamefont {W.}~\bibnamefont {Wester}}, \bibinfo {author} {\bibfnamefont
  {C.}~\bibnamefont {Boutan}}, \bibinfo {author} {\bibfnamefont {N.~S.}\
  \bibnamefont {Oblath}}, \bibinfo {author} {\bibfnamefont {R.}~\bibnamefont
  {Bradley}}, \bibinfo {author} {\bibfnamefont {E.~J.}\ \bibnamefont {Daw}},
  \bibinfo {author} {\bibfnamefont {A.~V.}\ \bibnamefont {Dixit}}, \bibinfo
  {author} {\bibfnamefont {J.}~\bibnamefont {Clarke}}, \bibinfo {author}
  {\bibfnamefont {S.~R.}\ \bibnamefont {O'Kelley}}, \bibinfo {author}
  {\bibfnamefont {N.}~\bibnamefont {Crisosto}}, \bibinfo {author}
  {\bibfnamefont {J.~R.}\ \bibnamefont {Gleason}}, \bibinfo {author}
  {\bibfnamefont {S.}~\bibnamefont {Jois}}, \bibinfo {author} {\bibfnamefont
  {P.}~\bibnamefont {Sikivie}}, \bibinfo {author} {\bibfnamefont
  {I.}~\bibnamefont {Stern}}, \bibinfo {author} {\bibfnamefont {N.~S.}\
  \bibnamefont {Sullivan}}, \bibinfo {author} {\bibfnamefont {D.~B.}\
  \bibnamefont {Tanner}}, \ and\ \bibinfo {author} {\bibfnamefont {G.~C.}\
  \bibnamefont {Hilton}} (\bibinfo {collaboration} {ADMX Collaboration}),\
  }\href {\doibase 10.1103/PhysRevLett.120.151301} {\bibfield  {journal}
  {\bibinfo  {journal} {Phys. Rev. Lett.}\ }\textbf {\bibinfo {volume} {120}},\
  \bibinfo {pages} {151301} (\bibinfo {year} {2018})}\BibitemShut {NoStop}%
\bibitem [{\citenamefont {Zhong}\ \emph {et~al.}(2018)\citenamefont {Zhong},
  \citenamefont {Al~Kenany}, \citenamefont {Backes}, \citenamefont {Brubaker},
  \citenamefont {Cahn}, \citenamefont {Carosi}, \citenamefont {Gurevich},
  \citenamefont {Kindel}, \citenamefont {Lamoreaux}, \citenamefont {Lehnert},
  \citenamefont {Lewis}, \citenamefont {Malnou}, \citenamefont {Maruyama},
  \citenamefont {Palken}, \citenamefont {Rapidis}, \citenamefont {Root},
  \citenamefont {Simanovskaia}, \citenamefont {Shokair}, \citenamefont
  {Speller}, \citenamefont {Urdinaran},\ and\ \citenamefont {van
  Bibber}}]{PhysRevD.97.092001}%
  \BibitemOpen
  \bibfield  {author} {\bibinfo {author} {\bibfnamefont {L.}~\bibnamefont
  {Zhong}}, \bibinfo {author} {\bibfnamefont {S.}~\bibnamefont {Al~Kenany}},
  \bibinfo {author} {\bibfnamefont {K.~M.}\ \bibnamefont {Backes}}, \bibinfo
  {author} {\bibfnamefont {B.~M.}\ \bibnamefont {Brubaker}}, \bibinfo {author}
  {\bibfnamefont {S.~B.}\ \bibnamefont {Cahn}}, \bibinfo {author}
  {\bibfnamefont {G.}~\bibnamefont {Carosi}}, \bibinfo {author} {\bibfnamefont
  {Y.~V.}\ \bibnamefont {Gurevich}}, \bibinfo {author} {\bibfnamefont {W.~F.}\
  \bibnamefont {Kindel}}, \bibinfo {author} {\bibfnamefont {S.~K.}\
  \bibnamefont {Lamoreaux}}, \bibinfo {author} {\bibfnamefont {K.~W.}\
  \bibnamefont {Lehnert}}, \bibinfo {author} {\bibfnamefont {S.~M.}\
  \bibnamefont {Lewis}}, \bibinfo {author} {\bibfnamefont {M.}~\bibnamefont
  {Malnou}}, \bibinfo {author} {\bibfnamefont {R.~H.}\ \bibnamefont
  {Maruyama}}, \bibinfo {author} {\bibfnamefont {D.~A.}\ \bibnamefont
  {Palken}}, \bibinfo {author} {\bibfnamefont {N.~M.}\ \bibnamefont {Rapidis}},
  \bibinfo {author} {\bibfnamefont {J.~R.}\ \bibnamefont {Root}}, \bibinfo
  {author} {\bibfnamefont {M.}~\bibnamefont {Simanovskaia}}, \bibinfo {author}
  {\bibfnamefont {T.~M.}\ \bibnamefont {Shokair}}, \bibinfo {author}
  {\bibfnamefont {D.~H.}\ \bibnamefont {Speller}}, \bibinfo {author}
  {\bibfnamefont {I.}~\bibnamefont {Urdinaran}}, \ and\ \bibinfo {author}
  {\bibfnamefont {K.~A.}\ \bibnamefont {van Bibber}},\ }\href {\doibase
  10.1103/PhysRevD.97.092001} {\bibfield  {journal} {\bibinfo  {journal} {Phys.
  Rev. D}\ }\textbf {\bibinfo {volume} {97}},\ \bibinfo {pages} {092001}
  (\bibinfo {year} {2018})}\BibitemShut {NoStop}%
\bibitem [{\citenamefont {Anastassopoulos}\ \emph {et~al.}(2017)\citenamefont
  {Anastassopoulos} \emph {et~al.}}]{cast2017}%
  \BibitemOpen
  \bibfield  {author} {\bibinfo {author} {\bibfnamefont {V.}~\bibnamefont
  {Anastassopoulos}} \emph {et~al.} (\bibinfo {collaboration} {CAST
  Collaboration}),\ }\href {http://dx.doi.org/10.1038/nphys4109} {\bibfield
  {journal} {\bibinfo  {journal} {Nature Physics}\ }\textbf {\bibinfo {volume}
  {13}},\ \bibinfo {pages} {584} (\bibinfo {year} {2017})}\BibitemShut
  {NoStop}%
\bibitem [{\citenamefont {Sikivie}\ \emph {et~al.}(2014)\citenamefont
  {Sikivie}, \citenamefont {Sullivan},\ and\ \citenamefont
  {Tanner}}]{PhysRevLett.112.131301}%
  \BibitemOpen
  \bibfield  {author} {\bibinfo {author} {\bibfnamefont {P.}~\bibnamefont
  {Sikivie}}, \bibinfo {author} {\bibfnamefont {N.}~\bibnamefont {Sullivan}}, \
  and\ \bibinfo {author} {\bibfnamefont {D.~B.}\ \bibnamefont {Tanner}},\
  }\href {\doibase 10.1103/PhysRevLett.112.131301} {\bibfield  {journal}
  {\bibinfo  {journal} {Phys. Rev. Lett.}\ }\textbf {\bibinfo {volume} {112}},\
  \bibinfo {pages} {131301} (\bibinfo {year} {2014})}\BibitemShut {NoStop}%
\bibitem [{\citenamefont {Sikivie}\ \emph {et~al.}(shed)\citenamefont
  {Sikivie}, \citenamefont {Sullivan},\ and\ \citenamefont {Tanner}}]{unp00s}%
  \BibitemOpen
  \bibfield  {author} {\bibinfo {author} {\bibfnamefont {P.}~\bibnamefont
  {Sikivie}}, \bibinfo {author} {\bibfnamefont {N.}~\bibnamefont {Sullivan}}, \
  and\ \bibinfo {author} {\bibfnamefont {D.~B.}\ \bibnamefont {Tanner}},\
  }\href@noop {} {} (\bibinfo {year} {unpublished}),\ \bibinfo {note}
  {unpublished work on the LC circuit axion dark matter detector was done in
  the early 2000's by the authors of ref.~\cite{PhysRevLett.112.131301} and
  independently by B. Cabrera and S.Thomas. The work of Cabrera and Thomas was
  presented in a talk.}\BibitemShut {Stop}%
\bibitem [{\citenamefont {Cabrera}\ and\ \citenamefont
  {Thomas}(shed)}]{Cabrera}%
  \BibitemOpen
  \bibfield  {author} {\bibinfo {author} {\bibfnamefont {B.}~\bibnamefont
  {Cabrera}}\ and\ \bibinfo {author} {\bibfnamefont {S.}~\bibnamefont
  {Thomas}},\ }\href
  {http://www.physics.rutgers.edu/~scthomas/talks/Axion-LC-Florida.pdf}
  {\enquote {\bibinfo {title} {Detecting string-scale qcd axion dark matter},}\
  } (\bibinfo {year} {unpublished}),\ \bibinfo {note} {2010 Axions Conference
  in Gainesville, Florida, January 15-17}\BibitemShut {NoStop}%
\bibitem [{\citenamefont {Ouellet}\ \emph {et~al.}(2019)\citenamefont
  {Ouellet}, \citenamefont {Salemi}, \citenamefont {Foster}, \citenamefont
  {Henning}, \citenamefont {Bogorad}, \citenamefont {Conrad}, \citenamefont
  {Formaggio}, \citenamefont {Kahn}, \citenamefont {Minervini}, \citenamefont
  {Radovinsky}, \citenamefont {Rodd}, \citenamefont {Safdi}, \citenamefont
  {Thaler}, \citenamefont {Winklehner},\ and\ \citenamefont
  {Winslow}}]{PhysRevLett.122.121802}%
  \BibitemOpen
  \bibfield  {author} {\bibinfo {author} {\bibfnamefont {J.~L.}\ \bibnamefont
  {Ouellet}}, \bibinfo {author} {\bibfnamefont {C.~P.}\ \bibnamefont {Salemi}},
  \bibinfo {author} {\bibfnamefont {J.~W.}\ \bibnamefont {Foster}}, \bibinfo
  {author} {\bibfnamefont {R.}~\bibnamefont {Henning}}, \bibinfo {author}
  {\bibfnamefont {Z.}~\bibnamefont {Bogorad}}, \bibinfo {author} {\bibfnamefont
  {J.~M.}\ \bibnamefont {Conrad}}, \bibinfo {author} {\bibfnamefont {J.~A.}\
  \bibnamefont {Formaggio}}, \bibinfo {author} {\bibfnamefont {Y.}~\bibnamefont
  {Kahn}}, \bibinfo {author} {\bibfnamefont {J.}~\bibnamefont {Minervini}},
  \bibinfo {author} {\bibfnamefont {A.}~\bibnamefont {Radovinsky}}, \bibinfo
  {author} {\bibfnamefont {N.~L.}\ \bibnamefont {Rodd}}, \bibinfo {author}
  {\bibfnamefont {B.~R.}\ \bibnamefont {Safdi}}, \bibinfo {author}
  {\bibfnamefont {J.}~\bibnamefont {Thaler}}, \bibinfo {author} {\bibfnamefont
  {D.}~\bibnamefont {Winklehner}}, \ and\ \bibinfo {author} {\bibfnamefont
  {L.}~\bibnamefont {Winslow}},\ }\href {\doibase
  10.1103/PhysRevLett.122.121802} {\bibfield  {journal} {\bibinfo  {journal}
  {Phys. Rev. Lett.}\ }\textbf {\bibinfo {volume} {122}},\ \bibinfo {pages}
  {121802} (\bibinfo {year} {2019})}\BibitemShut {NoStop}%
\bibitem [{\citenamefont {McAllister}\ \emph {et~al.}(2018)\citenamefont
  {McAllister}, \citenamefont {Goryachev}, \citenamefont {Bourhill},
  \citenamefont {Ivanov},\ and\ \citenamefont {Tobar}}]{beast_2018}%
  \BibitemOpen
  \bibfield  {author} {\bibinfo {author} {\bibfnamefont {B.~T.}\ \bibnamefont
  {McAllister}}, \bibinfo {author} {\bibfnamefont {M.}~\bibnamefont
  {Goryachev}}, \bibinfo {author} {\bibfnamefont {J.}~\bibnamefont {Bourhill}},
  \bibinfo {author} {\bibfnamefont {E.~N.}\ \bibnamefont {Ivanov}}, \ and\
  \bibinfo {author} {\bibfnamefont {M.~E.}\ \bibnamefont {Tobar}},\ }\href@noop
  {} {\  (\bibinfo {year} {2018})},\ \Eprint {http://arxiv.org/abs/1803.07755}
  {arXiv:1803.07755 [physics.ins-det]} \BibitemShut {NoStop}%
\bibitem [{\citenamefont {{Silva-Feaver}}\ \emph {et~al.}(2017)\citenamefont
  {{Silva-Feaver}}, \citenamefont {{Chaudhuri}}, \citenamefont {{Cho}},
  \citenamefont {{Dawson}}, \citenamefont {{Graham}}, \citenamefont {{Irwin}},
  \citenamefont {{Kuenstner}}, \citenamefont {{Li}}, \citenamefont {{Mardon}},
  \citenamefont {{Moseley}}, \citenamefont {{Mule}}, \citenamefont {{Phipps}},
  \citenamefont {{Rajendran}}, \citenamefont {{Steffen}},\ and\ \citenamefont
  {{Young}}}]{7750582}%
  \BibitemOpen
  \bibfield  {author} {\bibinfo {author} {\bibfnamefont {M.}~\bibnamefont
  {{Silva-Feaver}}}, \bibinfo {author} {\bibfnamefont {S.}~\bibnamefont
  {{Chaudhuri}}}, \bibinfo {author} {\bibfnamefont {H.}~\bibnamefont {{Cho}}},
  \bibinfo {author} {\bibfnamefont {C.}~\bibnamefont {{Dawson}}}, \bibinfo
  {author} {\bibfnamefont {P.}~\bibnamefont {{Graham}}}, \bibinfo {author}
  {\bibfnamefont {K.}~\bibnamefont {{Irwin}}}, \bibinfo {author} {\bibfnamefont
  {S.}~\bibnamefont {{Kuenstner}}}, \bibinfo {author} {\bibfnamefont
  {D.}~\bibnamefont {{Li}}}, \bibinfo {author} {\bibfnamefont {J.}~\bibnamefont
  {{Mardon}}}, \bibinfo {author} {\bibfnamefont {H.}~\bibnamefont {{Moseley}}},
  \bibinfo {author} {\bibfnamefont {R.}~\bibnamefont {{Mule}}}, \bibinfo
  {author} {\bibfnamefont {A.}~\bibnamefont {{Phipps}}}, \bibinfo {author}
  {\bibfnamefont {S.}~\bibnamefont {{Rajendran}}}, \bibinfo {author}
  {\bibfnamefont {Z.}~\bibnamefont {{Steffen}}}, \ and\ \bibinfo {author}
  {\bibfnamefont {B.}~\bibnamefont {{Young}}},\ }\href {\doibase
  10.1109/TASC.2016.2631425} {\bibfield  {journal} {\bibinfo  {journal} {IEEE
  Transactions on Applied Superconductivity}\ }\textbf {\bibinfo {volume}
  {27}},\ \bibinfo {pages} {1} (\bibinfo {year} {2017})}\BibitemShut {NoStop}%
\bibitem [{\citenamefont {Chu}\ \emph {et~al.}(2018)\citenamefont {Chu},
  \citenamefont {Duffy}, \citenamefont {Kim},\ and\ \citenamefont
  {Savukov}}]{PhysRevD.97.072011}%
  \BibitemOpen
  \bibfield  {author} {\bibinfo {author} {\bibfnamefont {P.-H.}\ \bibnamefont
  {Chu}}, \bibinfo {author} {\bibfnamefont {L.~D.}\ \bibnamefont {Duffy}},
  \bibinfo {author} {\bibfnamefont {Y.~J.}\ \bibnamefont {Kim}}, \ and\
  \bibinfo {author} {\bibfnamefont {I.~M.}\ \bibnamefont {Savukov}},\ }\href
  {\doibase 10.1103/PhysRevD.97.072011} {\bibfield  {journal} {\bibinfo
  {journal} {Phys. Rev. D}\ }\textbf {\bibinfo {volume} {97}},\ \bibinfo
  {pages} {072011} (\bibinfo {year} {2018})}\BibitemShut {NoStop}%
\bibitem [{\citenamefont {Crisosto}\ \emph {et~al.}(2018)\citenamefont
  {Crisosto}, \citenamefont {Sikivie}, \citenamefont {Sullivan},\ and\
  \citenamefont {Tanner}}]{10.1007/978-3-319-92726-8_15}%
  \BibitemOpen
  \bibfield  {author} {\bibinfo {author} {\bibfnamefont {N.}~\bibnamefont
  {Crisosto}}, \bibinfo {author} {\bibfnamefont {P.}~\bibnamefont {Sikivie}},
  \bibinfo {author} {\bibfnamefont {N.~S.}\ \bibnamefont {Sullivan}}, \ and\
  \bibinfo {author} {\bibfnamefont {D.~B.}\ \bibnamefont {Tanner}},\ }in\
  \href@noop {} {\emph {\bibinfo {booktitle} {Microwave Cavities and Detectors
  for Axion Research}}},\ \bibinfo {editor} {edited by\ \bibinfo {editor}
  {\bibfnamefont {G.}~\bibnamefont {Carosi}}, \bibinfo {editor} {\bibfnamefont
  {G.}~\bibnamefont {Rybka}}, \ and\ \bibinfo {editor} {\bibfnamefont
  {K.}~\bibnamefont {van Bibber}}}\ (\bibinfo  {publisher} {Springer
  International Publishing},\ \bibinfo {address} {Cham},\ \bibinfo {year}
  {2018})\ pp.\ \bibinfo {pages} {127--133}\BibitemShut {NoStop}%
\bibitem [{\citenamefont {Crisosto}(2018)}]{CrisostoThesis}%
  \BibitemOpen
  \bibfield  {author} {\bibinfo {author} {\bibfnamefont {N.}~\bibnamefont
  {Crisosto}},\ }\emph {\bibinfo {title} {Searching for Low Mass Axions with an
  LC Circuit}},\ \href@noop {} {Ph.D. thesis},\ \bibinfo  {school} {University
  of Florida} (\bibinfo {year} {2018})\BibitemShut {NoStop}%
\bibitem [{\citenamefont {Terman}(1950)}]{TermanEquation}%
  \BibitemOpen
  \bibfield  {author} {\bibinfo {author} {\bibfnamefont {F.}~\bibnamefont
  {Terman}},\ }\href@noop {} {\emph {\bibinfo {title} {Radio Engineers
  Handbook}}}\ (\bibinfo  {publisher} {McGraw-Hill},\ \bibinfo {year}
  {1950})\BibitemShut {NoStop}%
\bibitem [{Note1()}]{Note1}%
  \BibitemOpen
  \bibinfo {note} {Stahl HDC-50}\BibitemShut {NoStop}%
\bibitem [{Note2()}]{Note2}%
  \BibitemOpen
  \bibinfo {note} {Pasternak PE-SR405FL}\BibitemShut {NoStop}%
\bibitem [{Note3()}]{Note3}%
  \BibitemOpen
  \bibinfo {note} {Minicircuits ZX05-1+}\BibitemShut {NoStop}%
\bibitem [{\citenamefont {Friis}(1944)}]{friis1944}%
  \BibitemOpen
  \bibfield  {author} {\bibinfo {author} {\bibfnamefont {H.}~\bibnamefont
  {Friis}},\ }\href {\doibase 10.1109/JRPROC.1944.232049} {\bibfield  {journal}
  {\bibinfo  {journal} {Proceedings of the IRE}\ }\textbf {\bibinfo {volume}
  {32}},\ \bibinfo {pages} {419} (\bibinfo {year} {1944})}\BibitemShut
  {NoStop}%
\bibitem [{\citenamefont {Ulmer}\ \emph {et~al.}(2009)\citenamefont {Ulmer},
  \citenamefont {Kracke}, \citenamefont {Blaum}, \citenamefont {Kreim},
  \citenamefont {Mooser}, \citenamefont {Quint}, \citenamefont {Rodegheri},\
  and\ \citenamefont {Walz}}]{doi:10.1063/1.3271537}%
  \BibitemOpen
  \bibfield  {author} {\bibinfo {author} {\bibfnamefont {S.}~\bibnamefont
  {Ulmer}}, \bibinfo {author} {\bibfnamefont {H.}~\bibnamefont {Kracke}},
  \bibinfo {author} {\bibfnamefont {K.}~\bibnamefont {Blaum}}, \bibinfo
  {author} {\bibfnamefont {S.}~\bibnamefont {Kreim}}, \bibinfo {author}
  {\bibfnamefont {A.}~\bibnamefont {Mooser}}, \bibinfo {author} {\bibfnamefont
  {W.}~\bibnamefont {Quint}}, \bibinfo {author} {\bibfnamefont {C.~C.}\
  \bibnamefont {Rodegheri}}, \ and\ \bibinfo {author} {\bibfnamefont
  {J.}~\bibnamefont {Walz}},\ }\href {\doibase 10.1063/1.3271537} {\bibfield
  {journal} {\bibinfo  {journal} {Review of Scientific Instruments}\ }\textbf
  {\bibinfo {volume} {80}},\ \bibinfo {pages} {123302} (\bibinfo {year}
  {2009})},\ \Eprint {http://arxiv.org/abs/https://doi.org/10.1063/1.3271537}
  {https://doi.org/10.1063/1.3271537} \BibitemShut {NoStop}%
\bibitem [{\citenamefont {Ebrahimi}\ \emph {et~al.}(2016)\citenamefont
  {Ebrahimi}, \citenamefont {Stallkamp}, \citenamefont {Quint}, \citenamefont
  {Wiesel}, \citenamefont {Vogel}, \citenamefont {Martin},\ and\ \citenamefont
  {Birkl}}]{doi:10.1063/1.4958647}%
  \BibitemOpen
  \bibfield  {author} {\bibinfo {author} {\bibfnamefont {M.~S.}\ \bibnamefont
  {Ebrahimi}}, \bibinfo {author} {\bibfnamefont {N.}~\bibnamefont {Stallkamp}},
  \bibinfo {author} {\bibfnamefont {W.}~\bibnamefont {Quint}}, \bibinfo
  {author} {\bibfnamefont {M.}~\bibnamefont {Wiesel}}, \bibinfo {author}
  {\bibfnamefont {M.}~\bibnamefont {Vogel}}, \bibinfo {author} {\bibfnamefont
  {A.}~\bibnamefont {Martin}}, \ and\ \bibinfo {author} {\bibfnamefont
  {G.}~\bibnamefont {Birkl}},\ }\href {\doibase 10.1063/1.4958647} {\bibfield
  {journal} {\bibinfo  {journal} {Review of Scientific Instruments}\ }\textbf
  {\bibinfo {volume} {87}},\ \bibinfo {pages} {075110} (\bibinfo {year}
  {2016})},\ \Eprint
  {http://arxiv.org/abs/https://aip.scitation.org/doi/pdf/10.1063/1.4958647}
  {https://aip.scitation.org/doi/pdf/10.1063/1.4958647} \BibitemShut {NoStop}%
\bibitem [{\citenamefont {Dicke}(1946)}]{doi:10.1063/1.1770483}%
  \BibitemOpen
  \bibfield  {author} {\bibinfo {author} {\bibfnamefont {R.~H.}\ \bibnamefont
  {Dicke}},\ }\href {\doibase 10.1063/1.1770483} {\bibfield  {journal}
  {\bibinfo  {journal} {Rev. Sci. Inst.}\ }\textbf {\bibinfo {volume} {17}},\
  \bibinfo {pages} {268} (\bibinfo {year} {1946})}\BibitemShut {NoStop}%
\end{thebibliography}%



\end{document}